\documentclass[12pt]{article}

\usepackage{epsfig}
\usepackage{graphicx}

\font\smallrm = cmr8
\begin{document}

\baselineskip24pt

\centerline{\bf Quantum algorithm for nonlinear differential equations}

\bigskip
\centerline{Seth Lloyd,$^{1,2}$ Giacomo De Palma,$^{1,2}$ 
Can Gokler,$^3$ Bobak Kiani,$^{2,4}$}
\centerline{ Zi-Wen Liu,$^5$
Milad Marvian,$^6$ Felix Tennie,$^7$ Tim Palmer,$^7$}

{\smallrm
\centerline{1. Department of Mechanical Engineering, MIT,  
2. Research Lab for Electronics, MIT,}
\centerline{ 3. Engineering And Applied Sciences, 
Harvard University, 
4. Department of Electrical Engineering and Computer
Science, MIT,}
\centerline{ 5. Perimeter Institute, 6. Department of 
Electrical and Computer Engineering, Department of Physics, UNM}
\centerline{ 7. Department of Physics, 
University of Oxford}
\centerline{$^*$ to whom correspondence should be addressed: slloyd@mit.edu}
}

\bigskip\noindent{\it Abstract:} Quantum computers are known to provide
an exponential advantage over classical computers for the solution
of linear differential equations in high-dimensional spaces.    Here,
we present a quantum algorithm for the solution of {\it nonlinear} differential
equations.   The quantum algorithm provides an exponential advantage over 
classical algorithms for solving nonlinear differential equations.
    Potential applications include the Navier-Stokes equation,
plasma hydrodynamics, epidemiology, and more.

\vskip 1cm
Quantum computers have been shown to provide an exponential 
advantage over classical computers for the solution of linear differential
equations [1-2].   Such quantum linear differential equation solvers
provide a potential application for near term intermediate scale quantum
computers [3].     Many useful differential equations are nonlinear.
Previous efforts to develop quantum algorithms for nonlinear differential
equations resulted in methods that scaled poorly: the number of resources
required grew exponentially with the integration time [4].    
Here, we present
a quantum algorithm for the solution of nonlinear differential equations
where the number of resources required grows quadratically in 
the integration time, and logarithmically in the dimension of the
state space, thereby providing an exponential advantage over 
classical differential equation solvers.
The nonlinear differential equation algorithm considerably expands the 
set of potential applications of near term quantum computers.

The basic method that we employ is to encode the vector representing the
state of the system to be investigated as a quantum state.  As with
quantum linear differential equation solvers, the potential exponential
advantage over classical computers arises because the dimension of
that quantum state is exponential in the number of qubits/qudits
in the state, allowing the exploration of very high-dimensional state
spaces.    To implement the nonlinearity, we employ   
multiple copies of that state, allowing the solution of differential
equations whose nonlinear terms are polynomial in that state.   We combine
methods for simulating the dynamics of the nonlinear Schr\"odinger equation 
with quantum linear differential equation solvers to obtain an algorithm
that integrates the nonlinear equation over time $t$ using a number
of additional copies that scales quadratically in $t$ in regimes where the dynamics
of the original equation is `reasonably' stable -- i.e., the Lyapunov
exponents of the dynamics are not too large.  As with classical
methods for solving nonlinear equations, the accuracy of
the nonlinear equation solver depends on the underlying numerical
integration method used -- explicit, implicit, multi-time step, etc. -- and 
on the stability of the particular nonlinear equation to be integrated.  
The quadratic scaling of the algorithm presented here arises from
the intrinsic error scaling of the Euler forward
method and may be reduced by using more sophisticated numerical methods.

Let  $x \in {\cal C}^d$ be a state vector in a $d$-dimensional complex vector 
space.    We want to solve equations of the form
$${dx\over dt} + f(x) x = b(t), \eqno(1)$$
where $f(x)$ is a $d \times d$ matrix that is an order $m$ polynomial 
function of the vectors $x$ and $x^\dagger$.   In the supplementary
material, we show how to increase the dimensionality of $x$ by 1 to write       
$$f(x) = {x^\dagger}^{ \otimes m} F x^{\otimes m},\eqno(2)$$
for a suitable tensor $F$.    For the applications considered here,
we assume that $F$ is sparse and that its entries are readily computable.  
This will be the case, for example, for nonlinear equations 
which describe physical systems that are governed by local interactions. 

\bigskip\noindent{\it The nonlinear Schr\"odinger equation}

When $f$ is anti-Hermitian, and there is no driving term $b(t)$, we can solve
equations of the form (1) efficiently in the quantum mechanical
setting by implementing a nonlinear Schr\"odinger equation [5-11].   
The nonlinear Schr\"odinger equation for a single system arises by applying
the usual linear Schr\"odinger evolution to multiple identical interacting 
copies of the original system, and taking the
limit that the number of copies becomes large.   
See [5-11] and the Supplementary Material for details of
scaling and errors for this approach: because of its nonlinearity, many
questions about the domain of applicability of the nonlinear Schr\"odinger
equation remain open.

The basic approach is as follows.
Take $n>>m$ copies of the initial state  $x(0)^{\otimes n }$, and
apply the Hamiltonian
$$ H = -i  { n  \choose m}^{-1} \sum_{j_1\ldots j_m} 
F_{j_1 \ldots j_m}   
,\eqno(3)$$ 
to the inital state $x(0)^{\otimes n }$.   Here $F_{j_1\ldots j_m}$
is the tensor $F$ applied to the $m$ distinct
subsystems labeled by
$j_1 \ldots j_m$.    
Look at the short-time behavior
of the $n$-system linear Schr\"odinger equation:
$$e^{-iH\Delta t} x^{\otimes n} 
= (I - iH\Delta t - (1/2) H^2 \Delta t^2 + O(\Delta t^3) ) x^{\otimes n}.
\eqno(4)$$
The short time behavior of any one of the copies is obtained by tracing
out the other copies, and one obtains the effective single system dynamics
$$ x \rightarrow ( I -  \Delta t f(x)) x + O(E^2\Delta t^2) ,\eqno(5)$$
where $E$ is the average value of $|f(x)|$ over that time period.  
That is, the short time behavior of each copy obeys equation (1) for
no driving. 
The deviations from the correct nonlinear behavior are suppressed by
a factor of $1/n$ in the second and higher order terms (see [5-11]
and Supplementary Material).

Equation (5) shows that to integrate the nonlinear Schr\"odinger 
equation over $T = t/\Delta t$ discretized Trotter steps, 
we need to take $T$ sufficiently large that $ E^2 T\Delta t^2 = E^2 t \Delta t  
= E^2 t^2/T$ is small.   That is, the number of discretized steps
required grows quadratically with the time over which the equation
is to be integrated.  In the Supplementary Material, we   
compare the discretized integration of the full multi-system time
evolution, equation (4), with the discretized integration 
of the nonlinear single system evolution, equation (5), and count terms
that differ between the two integrations at each order in $t$.  
We show that as long as the number of copies is significantly greater than $T$,
the nonlinear Schr\"odinger equation approximation holds to accuracy
$\epsilon$ for times $t$ such
that $E^2 t \Delta t m^2/n < \epsilon$.   That is, 
the number of copies required also scales quadratically in $t$. 

In addition, we have to pay attention to the tendency
of the nonlinear Schr\"odinger equation to amplify small deviations
in the wave function exponentially: the accuracy of the numerical
integration of the  
nonlinear Schr\"odinger equation breaks down when the nonlinearity
amplifies the energy $E$ sufficiently that 
$E^2 t\Delta t > O(\epsilon)$. 

The nonlinear Schr\"odinger equation is a mean-field equation
in which the nonlinear unitary dynamics of a single system is
determined by weak interactions with many other systems in the same state.
Deviation from the single-system unitary nonlinear
Schr\"odinger dynamics come from entangling terms in the dynamics, which
only arise at second order in equation (4), and which are suppressed
by a factor of $1/n$.   In practice, the nonlinear Schr\"odinger equation
can provide a highly accurate description of weakly interacting
identical particles: a well-known example of its application is the
Gross-Pitaevskii equation which describes the dynamics of bosons in
a Bose-Einstein condensate, where each boson can be described by the
same wave function.   In the case of the Gross-Pitaevskii equation,
the the tensor $F$ is sparse: its linear part is simply the
single-particle Schr\"odinger equation, and its nonlinear
part represents a spatially local interaction between the bosons.

\bigskip\noindent{\it Quantum solution to general nonlinear differential equations}

As just seen, the case of a nonlinear equation where $f(x)$ is anti-Hermitian 
is a special case 
that is tailor-made for solution by quantum
computers: simply use standard techniques of quantum simulation to simulate
the dynamics induced by the Hamiltonian of equation (3).  Equation (1)
without driving then takes the form of a nonlinear Schr\"odinger equation, and
the resulting multi-system quantum dynamics goes through a series
of quantum states that represent the solution to equation (1).
Features of the
solution can be obtained by performing quantum post-processing on the
quantum states generated by the simulation [3].

It is not particularly surprising that we can use a quantum computer
to solve a nonlinear Schr\"odinger equation: we simply perform a quantum
simulation of the symmetric multi-system quantum dynamics that leads
to such equations.   The case of general nonlinear differential equations 
is harder.   When $f(x)$ is not anti-Hermitian, equation
(1) encompasses a wide variety of nonlinear differential equations, including
the Boltzmann equation, the Navier-Stokes equation, plasma hydrodynamics, etc.
Since $f(x)$ is not anti-Hermitian, we can't embed its solution into a 
nonlinear Schr\"odinger equation.   We now show, however, that we can
combine the mean-field techniques of the nonlinear Schr\"odinger equation with
the methods of quantum linear differential equation solvers [1-2] to obtain
a quantum solution to equation (1).    The resulting quantum nonlinear
differential equation algorithm inherits the exponential quantum advantage
of the linear differential equation solvers.

\bigskip\noindent{\it Review of quantum linear differential equation solvers}

To set up the general nonlinear case, we first review the methods by
which quantum algorithms obtain a quantum
representation of the solution of linear differential equations.   To
obtain the optimal scaling and computational complexity, sophisticated
methods are required [2].   For the sake of simplicity of exposition,
we review here the original method proposed in [1], mentioning
extensions and elaborations as we go.

[1] showed how to map the problem of solving a general linear differential
equation to that of matrix inversion, which can then be performed using
the quantum linear systems algorithm [12-13].   Consider a linear differential
equation of the form,
$$ {dx\over dt} + A x = b(t), \eqno(6)$$
where as above $x,b \in {\cal C}^d$ and $A$ is a $d\times d$ matrix.
Discretize the equation in time at intervals $\Delta t$, and take $k$ to
be the index for the discretized time, so that $x_k$ and $b_k$ are
the values of $x$ and $b$ at time label $k$.   We wish to integrate equation
(6) numerically starting from the initial state $x_0 \equiv b_0$.   We obtain
a series of equations of the form: 
$$ 
 x_0 = b_0 \quad 
x_1 = x_0 - \Delta t A x_0 + \Delta t b_1 \quad 
\ldots \quad
x_{k+1} = x_k - \Delta t A x_k + \Delta t  b_k \quad 
\ldots 
\eqno(7)$$ 
Here, we have used the Euler forward method for numerical integration, but it 
is straightforward to implement implicit methods such as
Euler backward, Crank-Nicholson, Runge-Kutta, etc., if greater numerical stability is
required [3].   
Written in matrix form, these equations become
$$- \pmatrix{- I & 0 & 0 & \ldots & 0 & 0 \cr
I-\Delta t A & - I & 0 & \ldots & 0 & 0 \cr
0 & I-\Delta t A& -I &  \ldots & 0 & 0\cr
&&\ldots &&&\cr
0& 0 & 0 & \ldots & - I & 0\cr
0& 0 & 0 & \ldots & I - \Delta t A & - I\cr    }
\pmatrix{ x_0 \cr x_1 \cr x_2 \cr \ldots \cr x_{T-1} \cr x_{T} \cr}
=  
\pmatrix{ b_0 \cr \Delta t b_1 \cr \Delta t b_2 \cr \ldots \cr \Delta t 
b_{T-1} \cr \Delta t b_{T} \cr}. \eqno(8)
$$

Writing in quantum form, we adjoin a time-step register $|k\rangle$,
and encode the solution vector whose components are the $x_k$, 
and the initial state/driving 
vector whose components are $b_k$, 
as (unnormalized) quantum `history states'
$$  |X\rangle = \sum_k |x_k\rangle |k\rangle, \quad  
|B\rangle = |b_0\rangle|k=0\rangle + \Delta t 
\sum_{k=1}^T |b_k\rangle |k\rangle.
\eqno(9)$$   
Here $|x_k\rangle$ is the quantum state corresponding to the
 solution to equation (8) at time-step $k$.
When $b_j =0$ for $j>0$ (no driving), the states $|x_k\rangle$
take the form
$$|x_k\rangle = 
\sum_k (I-\Delta t A)^k |x_0\rangle \approx 
\sum_k e^{-k\Delta t A} |x_0\rangle . \eqno(10)$$  
The matrix in equation (8) can be written in quantum form as
$${\cal M} =  \sum_{k=0}^T I\otimes |k\rangle\langle k|
- \sum_{k=0}^{T-1} (I - \Delta t A) \otimes |k+1\rangle\langle k|.\eqno(11)$$ 

We now use the quantum linear system algorithm [12-13] to solve the equation
$${\cal M} |X\rangle =  |B\rangle. \eqno(12)$$
The quantum algorithm takes as inputs the matrix $A$, the initial state,
and the vector of driving terms, and returns as its solution
the history state $|X\rangle$,
revealing its normalization in the process.   This quantum history
state can now be measured and undergo quantum post-processing [3] to reveal
features of the solution to equation (6).

\bigskip\noindent{\it Non-Hermitian nonlinear Schr\"odinger equation}

We now combine quantum linear differential equation solvers with
techniques borrowed from the treatment of the nonlinear Schr\"odinger
equation [5-11] to implement a quantum nonlinear differential equation solver.
The central insight is that the construction of the quantum history
state equation (9) via the quantum linear differential equation
solver does not require the matrix $A$ to be Hermitian.
So we simply implement a quantum version of the dynamics of equation
(1) by constructing a nonlinear Schr\"odinger equation for a
non-Hermitian Hamiltonian, and use the quantum linear systems algorithm
to construct the desired history state that corresponds to the
solution of equation (1).  Just as in the Hermitian case, the non-Hermitian
case preserves the tensor product form of the solution state
up to terms which are suppressed by $O(1/n)$. 

Explicitly: Take $n$ copies of the input/driving state and construct
the state
$$ |B\rangle^{ (n)} \equiv |b_0\rangle^{\otimes n} |k=0\rangle
+   \Delta t \sum_{k=1}^T |b_k\rangle^{\otimes n} 
|k\rangle. \eqno(13)$$
For the matrix ${\cal M}$, we use the operator
$${\cal M}^{(n)} \equiv  \sum_{k = 0}^T  I \otimes |k\rangle\langle k|
- \sum_{k = 0}^{T-1} (I - \Delta t { n  \choose m}^{-1} \sum_{j_1\ldots j_m}
F_{j_1 \ldots j_m}) \otimes |k+1\rangle\langle k|.\eqno(14)$$
The solution to the equation
$$ {\cal M}^{(n)} |X\rangle^{(n)} = |B\rangle^{(n)},\eqno(15)$$
then takes the form of the history state
$$|X\rangle^{(n)} \approx 
\sum_k |x^1_k\rangle \otimes \ldots \otimes |x^n_k\rangle |k\rangle,
\eqno(16)$$
where
$$|x_k\rangle =     (I - \Delta t f(x_{k-1})) |x_{k-1}\rangle + 
\Delta t |b_k\rangle. \eqno(17)$$
$|x_k\rangle$ is the Euler forward solution to equation (1) at
time $k$, 
and $|x^j_k\rangle = |x_k\rangle$ for all $j$.  
The approximation sign in equation (16) comes from the error
induced by the nonlinear Schr\"odinger equation approximation
(see Supplementary Material), and  means that 
the marginal density matrices for each subsystem at
time-step $k$ are equal to
$$|x_k\rangle\langle x_k| + O(|E|^2 T m^2\Delta t^2/n). \eqno(18) $$ 
Here, $|E|^2$ is the average
modulus squared for the complex eigenvalues of the non-Hermitian 
Hamiltonian $f(x)$ over the integration time.  
That is, the $n$-fold tensor product history state is, to lowest
order in $1/n$, the superposition
of the $n$-fold tensor product of the solutions to the desired nonlinear
equation (1), at different times.
As with the Hermitian nonlinear Schr\"odinger equation, 
the error is dominated by discretization error of the numerical
integration method, and we require that
$$ \quad |E|^2 t \Delta t < 
O(\epsilon)\eqno(19)$$
throughout the integration time $t$.   
As shown in the supplementary material,
the error due to the nonlinear Schr\"odinger equation approximation is
$|E|^2 t\Delta t m^2/n$, which is suppressed by a factor of $n$
compared with the error of the numerical integration.

\bigskip\noindent{\it Errors, stability, and range of applicability}

For ease of explication, we have presented the simplest possible version
of the quantum nonlinear differential equation algorithm, which is based
on lowest order Trotterization [14].    Higher order Trotterization, and various
other techniques such as the use of higher order implicit numerical methods,
are likely to improve the error scaling.   The approximations
leading to the quantum solution, equation (17) of the nonlinear equation
(1), must necessarily break down if the nonlinearity leads to large
exponential growth, e.g., if the nonlinear equation has positive
Lyapunov exponents, and we try to integrate for times sufficiently long
that equation (19) is violated.    Indeed, if they did not break down, one could
use the method to amplify exponentially small differences in the initial
wave function, which would allow the solution of NP-complete problems 
on a quantum computer [15-17]!    Note that this issue also arises
for the quantum solution of linear equations when the governing matrix has
eigenvalues with positive real part.    In both the linear and the nonlinear
cases, the quantum solution gives an exponential speed up only over times
where such amplification does not result in violations of equation
(19).  

The long time accuracy
of the quantum nonlinear equation solver in the presence of
positive Lyapunov exponents can be better than that of the quantum
linear equation solver with positive real eigenvalues, because
the nonlinearity implies that the directions in which the solution
is exponentially expanding change over time, so that there is
no net exponential growth: this is the case, for example, when the sum
of the Lyapunov exponents is negative, leading to fractal solutions
such as strange attractors.    In the nonlinear case,
because of its construction, the quantum solver 
works best when applied to problems where the mean-field approximation
remains valid throughout the period of the time evolution that
is investigated, and where the underlying numerical method (Euler forward,
Euler backward, Crank-Nicholson, Runge-Kutta, etc.) is accurate.   
The performance of the quantum nonlinear
differential equation algorithm will vary depending on the characteristics
-- notably, the stability -- of the nonlinear equation that it is 
given to solve.    

\bigskip\noindent{\it Applications:}

To obtain the exponential speed-up afforded by
the quantum nonlinear differential equation algorithm,
the tensor $F$ must be sparse and have computable entries.
In the case of the Boltzmann equation, 
for example, the vector $x$ represents a vector of probabilities or
densities $p(y,v)$ in the single-particle phase space of positions 
$y$ and velocities $v$, the linear
part of the tensor $F$ represents the noninteracting diffusive dynamics
of the fluid, and the nonlinear part of $F$ represents a spatially
local, momentum conserving scattering dynamics.  The spatial
locality of the interaction, combined with the well-specified
form of the scattering interaction, implies that $F$ is sparse
and that its entries are computable. 
Similarly, in the Navier-Stokes equation, the locality of
interactions combined with momentum conservation gives rise
to sparse, computable tensors $F$.   The nonlinear equations
for plasma hydrodynamics include both the dynamics of charged
particles and of the electromagnetic fields: although these
equations are more complex, because they involve local interactions
and dynamics that obey the laws of physics the resulting nonlinear terms
in $F$ are again sparse and computable. 

\bigskip\noindent{\it Comparison to related work:} 

The first effort to construct a quantum algorithm for nonlinear
differential equations [4] also used multiple copies of the system
to induce the nonlinearity, but because the algorithm required
that $m$ copies at each Trotter step be sacrificed
to produce a single copy at the
next step, the resources required by the algorithm scaled as
exponentially in the number of Trotter steps $T$: 
for pairwise nonlinear interactions, for example, the resources required
in [4] scale as $2^T$.   
The nonlinear, non-Hermitian Schr\"odinger equation
approach presented here obtains quadratic scaling in $T$ by retaining
all copies of the system at each step, rather than discarding them.
Other efforts to present quantum algorithms for nonlinear
differential equations rely on variational techniques [18], 
but do not supply the provable exponential speed-up given by our
algorithm.   A separate method
involves the Madelung hydrodynamic approach to 
quantum mechanics [19].
A recently posted work [20] presents a method similar
to that pursued here, using a linear system over
multiple copies to induce the single-system nonlinearity and
applying the quantum linear differential equation solver,
but uses classical Carleman linearization instead of
the quantum nonlinear Schr\"odinger linearization technique. 

\bigskip\noindent{\it Conclusion:}

This paper showed that quantum computers can in principle attain an
exponential advantage over classical computers for solving nonlinear
differential equations.
The main potential advantage of the quantum nonlinear
equation algorithm over classical algorithms is that it scales logarithmically
in the dimension of the solution space, making it a natural candidate
for applying to high dimensional problems such as 
the Navier-Stokes equation and other nonlinear fluids, plasmas, etc.
The method developed here could be applied to continuous variable
quantum systems using the techniques of [21].
Like quantum linear differential equation solvers, 
the quantum nonlinear solver presents its solution as a quantum 
history state, equation (9), which allows quantum post-processing [3] to
extract features such as the power spectrum (quantum Fourier transform), 
principal components (quantum singular value transformation), and
multiscale behavior (quantum wavelet transforms).

\vskip 1cm

\bigskip\noindent{\it Acknowledgements:} 
This work was supported by DOE, ARO, AFOSR, and DARPA.  BK was 
supported by an MITEI Exxon Fellowship.
TP was supported by a Royal Society Research Professorship and FT by a Royal Society Research Professorship Enhancement Award. 
ZWL is supported by Perimeter Institute for Theoretical Physics.
Research at Perimeter Institute is supported in part by the Government of Canada through the Department of Innovation, Science and Economic Development Canada and by the Province of Ontario through the Ministry of Colleges and Universities.

\vskip 1cm

\noindent{\it References:}

\noindent [1]  D.W. Berry,  {\it J. Phys. A: Math. Theor. } {\bf 47} 105301 (2014);
 arXiv: 1010.2745.

\noindent [2] D.W. Berry, A.M. Childs, A. Ostrander, G. Wang, 
{\it Comm. Math. Phys.} {\bf 356}  1057–1081 (2017); arXiv: 1701.03684.

\noindent [3] B.T. Kiani, G. De Palma, D. Englund, W. Kaminsky,
M. Marvian, S. Lloyd, 
`Quantum advantage for differential equation analysis,'
arXiv: 2010.15776 (2020).

\noindent [4] S.K. Leyton, T.J. Osborne, `A quantum algorithm 
to solve nonlinear differential equations,' arXiv: 0812.4423 (2008).

\noindent [5] K. Hepp, {\it Comm. Math. Phys.} {\bf 35}(4), 265–277 (1974).

\noindent [6] H. Spohn, {\it Rev. Mod. Phys.} {\bf 52}, 569–615 (1980).

\noindent [7] I. Rodnianski, B. Schlein, {\it Comm. Math. Phys.} {\bf 291},
31–61 (2009).

\noindent [8] L. Erd\"os, B. Schlein, {\it J. Stat. Phys.} {\bf 134}, 
859–870 (2009).

\noindent [9] L. Chen, J.O. Lee, B. Schlein, {\it J. Stat. Phys.}
{\bf 144}(4), 872 (2011).

\noindent [10] N. Benedikter, M. Porta, B. Schlein, {\it  
Effective Evolution Equations from Quantum Dynamics,} Springer 
International (2015).

\noindent [11] C. Gokler, `Mean field limit for many-particle interactions,' 
arXiv: 2006.05486 (2020).

\noindent [12] A.W. Harrow, A. Hassidim, S. Lloyd, {\it Phys. Rev. Lett.}
{\bf 103}, 150502 (2009); arXiv: 0811.3171. 

\noindent [13] A.M. Childs, R. Kothari, R.D. Somma, {\it SIAM J. Comp.}
{\bf 46}  1920–1950 (2017); arXiv: 1511.02306.

\noindent [14] S. Lloyd, {\it Science} {\bf 273}, 1073-1078 (1996).

\noindent [15] D.S. Abrams, S. Lloyd, {\it Phys. Rev. Lett.} {\bf 81} 
3992 (1998); arXiv: quant-ph/9801041.

\noindent [16] S. Aaronson, {\it ACM SIGACT News} {\bf 36}  30–52 (2005);
arXiv: quant-ph/0502072.

\noindent [17] A.M. Childs, J. Young, {\it Phys. Rev. A}
{\bf 93}  022314 (2016); arXiv: 1507.06334.

\noindent [18] M. Lubasch, J. Joo, P. Moinier, M. Kiffner, D. Jaksch, 
{\it Phys. Rev. A} {\bf 101}, 010301(R) (2020).

\noindent [19] I.Y. Dodin and E.A. Startsev, 
`On applications of quantum computing to plasma 
simulations, arXiv: 2005.14369 (2020).

\noindent [20] J.-P. Liu, H.O. Kolden, H.K. Krovi, N.F. Loureiro,
K. Trivisa, A.M. Childs, `Efficient quantum algorithm for
dissipative nonlinear differential equations, 
arXiv: 2011.03185.

\noindent [21] J.M. Arrazola, T. Kalajdzievski, C. Weedbrook, S. Lloyd,
{\it Phys. Rev. A} {\bf 100}, 032306 (2019).

\vskip 1cm
\noindent {\it Supplementary material:}

We show here how to write an arbitary $m$'th order polynomial
in the form of equation (2):
$$f(x) = {x^\dagger}^{ \otimes m} F x^{\otimes m}.\eqno(S1)$$
Let $x$ be $d$-dimensional: 
$$x = (x_1, \ldots, x_d)^{\rm T}.\eqno(S2a)$$
Add an extra dimension:
$$\tilde x = (x_0, x_1, \ldots, x_d)^{\rm T},\eqno(S2b)$$
and augment the original differential equation (1) with the
initial condition $x_0 = 1$, and $dx_0/dt = 0$, so that
$x_0$ remains 1 at all times.   

The addition of an extra dimension whose coefficient takes 
a constant value allows us to include arbitrary polynomials
in the $\{ x_j, \bar x_j \}$ in $f(x)$. 
For example, in the case $m = 2$, to include the monomial $\bar x_1^2 x_2$ in
the $|2\rangle \langle 1|$ entry of $f(x)$ (here we use
quantum notation), 
we include in $F$ a term 
$$ G_{\bar 1 \bar 1 2} = |2\rangle \langle 1| \otimes(1/2)  
(|1\rangle_1 \langle 2| 
\otimes |1\rangle_2 \langle 0| 
+ |1\rangle_2 \langle 0| 
\otimes |1\rangle_2 \langle 2|).\eqno(S3)$$
Performing the inner products yields:
$$ \langle x|\langle x| G_{\bar 1 \bar 1 2} |x\rangle |x\rangle
= \bar x_1^2 x_2 |2\rangle \langle 1|, \eqno(S4)$$
the desired result.

\bigskip\noindent{\it Normalization}

In the usual, Hermitian version of the nonlinear Schr\"odinger equation,
the states $|x_j\rangle$ automatically remain normalized to one.
To maintain this normalization under the action of 
the non-Hermitian operator of equation (14), 
we rescale the variables so that $\tilde x^\dagger \tilde x \leq 1/2$
for the period over which we integrate the equation, and   
add an additional dimension whose real coefficient $x_{d+1}$ obeys
$x_{d+1}^2 = 1- \tilde x^\dagger \tilde x $.   To maintain the 
normalization throughout the integration period, $x_{d+1}$ is taken
to obey the nonlinear equation
$$ {dx_{d+1} \over dt} = {d\over dt} (1- \tilde x^\dagger \tilde x)^{1/2}
=   -(1/2) (1- \tilde x^\dagger \tilde x)^{-1/2} 
\big( {d\tilde x^\dagger \over dt} 
\tilde x + \tilde x^\dagger {d\tilde x \over dt} \big) \eqno(S5),$$
which we implement within our framework by expanding 
$(1- y)^{-1/2}$ as a Taylor series about the point $y=0$.
The Taylor series converges exponentially, and
so we augment $m$ by $\log(1/\epsilon)$ to
maintain accuracy $\epsilon$ throughout.

\bigskip\noindent {\it Accuracy of the nonlinear Schr\"odinger equation}

We compare the time evolution of the full multi-system linear 
Schr\"odinger equation with the nonlinear Schr\"odinger equation.
The analysis holds for both the Hermitian and non-Hermitian case.
Because the algorithm works by discretization in time, 
we write the discretized Schr\"odinger evolution for the full system as 
$$e^{-iT\Delta t H} |x_0\rangle^{\otimes n} 
\approx (I - i \Delta t H)^T |x_0\rangle^{\otimes n} = \bigg( I - 
\Delta t {n \choose m}^{-1} \sum_{i_1 \ldots i_m} F_{i_1\ldots i_m} \bigg)^T |x_0\rangle^{\otimes n}, \eqno(S6)$$
and the accuracy holds to the usual first order discretized
approximation [16]. 
%the error goes as 
%$${n \choose m}^{-2} \sum_{i_1 \ldots i_m,i'_1 \ldots i'_m} 
%[ F_{i_1\ldots i_m}, F_{i'_1\ldots i'_m}] t\Delta t.  \eqno(S6)$$
We compare this with the discretized evolution of a single copy
under the nonlinear Schr\"odinger
equation:
$$(1-f(x_{T-1})\Delta t) (1-f(x_{T-2})\Delta t ) \ldots
(1-f(x_0)\Delta t) |x_0\rangle. \eqno(S7)$$

Rewrite (S6) in density matrix form:
$$\bigg( I -
\Delta t {n \choose m}^{-1} \sum_{i_1 \ldots i_m} F_{i_1\ldots i_m} \bigg)^T
(|x_0\rangle\langle x_0|)^{\otimes n}
\bigg( I -
\Delta t {n \choose m}^{-1} \sum_{i'_1 \ldots i'_m} F^\dagger_{i'_1\ldots i'_m} 
\bigg)^T
. \eqno(S8)$$
First, look at single step: $T=1$.   The time evolution of the
first subsystem is obtained by tracing out 
subsystems $2 \ldots n$ in equation (S8).
The first order terms are the same as in the discretized
nonlinear Schr\"odinger dynamics.    At second order,
terms where the set of indices $i_1 \ldots i_m$ have no overlap with 
the set of indices $i'_1\ldots i'_m$ give a dynamics which is exactly
the discretized nonlinear Schr\"odinger equation dynamics:
$$|x_0\rangle\langle x_0| \rightarrow |x_1\rangle\langle x_1| = 
(1-f(x_0)\Delta t) |x_0\rangle\langle x_0| (1-f^\dagger(x_0)\Delta t).
\eqno(S9)$$
Second order terms that do have overlap between the two sets of indices
represent potentially entangling dynamics that departs from the 
tensor product form of the solution.
The fraction of terms in which no such overlap occurs is approximately 
equal to, and bounded above by 
$$(1-m/n)^m \approx 1-m^2/n.\eqno(S10)$$
This approximation holds as long as $n>>mT$: that is, the number
of copies must be larger than the number of time steps $T$.
That is, the state produced by the full dynamics $(S6)$ yields
an overall state that is approximately the desired state,
$$ |x_1\rangle\langle x_1|^{\otimes n},\eqno(S11)$$ 
with single-system density matrices that are of the form 
$$|x_1\rangle\langle x_1| + O(|E|^2 \Delta t^2 m^2/n),\eqno(S12)$$
where we have included the maximum energy and time scales to give
the size of the deviation from the correct nonlinear Schr\"odinger
dynamics at that step.   Because the errors are generated by entangling
each subsystem with all other subsystems, the error terms for any 
subset of subsystems manifest themselves as
mixed states, so the state of
any subset of $m$ subsystems together with the errors is the
state
$$ (1-\epsilon_1)|x_1\rangle\langle x_1|^{\otimes m} + \epsilon_1 \eta_1  
\eqno(S13)$$
where $\epsilon_1$ is $O(|E|^2 \Delta t^2 m^2/n)$ 
and $\|\eta_1\|_1 \leq 3$. 
For the $m$-fold density matrix, the fraction of
terms in (S8) that give departures from the tensor product form is
approximately $2m^2/n$, because we have to also count terms that entangle the
$m$ subsystems with each other.  
This form for the $m$-fold reduced density matrices means that
at the next step, the errors induced by the presence of entanglement
and correlations are additive.

Now apply the second time step: exactly the same argument applies,
and we obtain a state 
whose single-system reduced density matrices
are of the form 
$$|x_2\rangle\langle x_2| + O(2 \overline{|E|^2} \Delta t^2 m^2/n).\eqno(S14)$$
where $\overline{|E|^2}$  is the average modulus squared energy scale
over the two steps (the energy scale at the 
second step can differ from the first).
The $m$ system reduced density matrices, for the same reason as before,
are of the form
$$(1- \epsilon_2)|x_2\rangle\langle x_2|^{\otimes m} + \epsilon_2 \eta_2,
\eqno(S15)$$
where $\epsilon_2$ is $O(2\overline{|E|^2 } \Delta t^2 m^2/n)$ 
and $\|\eta_2\|_1\leq 3$.  
Continuing for $T$ steps yields a final state whose single-system
density matrices are of the form 
$$|x_T\rangle\langle x_T| + O(  \overline{|E|^2} T \Delta t^2 m^2/n).\eqno(S16)$$
That is, for small $m$ the errors introduced by the 
nonlinear Schr\"odinger equation approximation are suppressed 
by a factor of $n$ compared with 
the discretization errors for the first order Trotterization of
the Euler forward method.  

\vfill\eject

\end{document}